\documentclass[apjl]{emulateapj}
\begin{document}
\title{Normalization of the Matter Power Spectrum via the Ellipticity 
Function of Giant Galaxy Voids from SDSS DR5}
\author{Jounghun Lee}
\affil{Department of Physics and Astronomy, FPRD, Seoul National University, 
Seoul 151-747, Korea: jounghun@astro.snu.ac.kr}
\begin{abstract}
The ellipticity function of cosmic voids exhibits strong dependence on the 
amplitude of the linear matter power spectrum. Analyzing the most recent 
void catalogs constructed by Foster and Nelson from the fifth data release 
of the Sloan Digital Sky Survey, we measure observationally the ellipticity 
function of giant galaxy voids. Then, we incorporate the redshift distortion 
and galaxy bias effect into the analytic model of the void ellipticity 
function and fit it to the observational result by adjusting the value of 
the power-spectrum normalization with the help of the generalized 
$\chi^{2}$-minimization method. The best-fit normalization of the linear 
power spectrum is found to be $\sigma_{8}=0.90\pm 0.04$. Our result is higher 
than the WMAP $\sigma_{8}$-value but consistent with that from the recent 
work of Liu and Li who have constructed a new improved CMB map independently.
\end{abstract}
\keywords{cosmology:theory --- large-scale structure of universe}

\section{INTRODUCTION}
 
The normalization amplitude of the linear matter power spectrum is one of the 
key cosmological parameters that are required to complete the theoretical 
description of the initial conditions of the universe \citep{teg-etal06}. 
It is often quantified in terms of $\sigma_{8}$, the rms fluctuations of the 
linear density field within a top-hat spherical radius $8\, h^{-1}$Mpc. 
Various observables have so far been used to constrain the value of 
$\sigma_{8}$: the cluster abundance \citep[e.g.,][]{hen-etal09}, the weak 
lensing cosmic shear \citep[e.g.,][]{wae-etal00}, strong lensing arc 
statistics \citep[e.g.,][]{HM04}, the cluster shapes \citep{lee06}, 
and the cosmic microwave background radiation (CMB) temperature map 
\citep[e.g.,][]{wmap5,LL09}. Yet, these observables depend not solely 
on $\sigma_{8}$ but concurrently on the other key parameters such as the 
matter density parameter $\Omega_{m}$, primordial non-Gaussianity 
parameter $f_{\rm NL}$, and dark energy equation of state $w$. 
Furthermore, it has been realized that complicated systematics involved 
in the measurement of these observables could bias strongly the 
estimates of $\sigma_{8}$. Hence, to break the parameter degeneracy and to 
diminish any systematic bias, it is very important to consider as many 
alternative probes as possible.

Recently, \cite[][hereafter, PL07]{PL07} have proposed the void ellipticity 
function as another sensitive probe of $\sigma_{8}$. Noting that the shapes 
of voids are modulated by the competition between tidal distortion and 
cosmic expansion, they have analytically derived the void ellipticity 
function under the assumption that the dynamics of void galaxies can be 
well described by the Zel'dovich approximation just as that of the dark 
matter particles are in the linear regime. They have tested their model 
against the results from the Millennium Run simulations \citep{spr-etal05}, 
which proved the validity of the PL07 model. For the comparison with 
observational data, however, the PL07 model has to be extended to 
incorporate the redshift distortion effect since in pratice the void 
ellipticities can be measured only in redshift space.

Moreover, there is one condition that the success of the PL07 analytic model 
is contingent upon. Its validity has been tested only for the case that the 
voids are found through the specific void-finding algorithm of 
\citet[][hereafter HV02]{HV02}. 
Since there is no unique definition of voids, the ellipticity distribution 
may well depend on the way in which voids are identified \citep{col-etal08}.
For the fair comparison with the PL07 model, the HV02 algorithm should be 
consistently used for the identification of voids from observations. 
Very recently, \citet[][hereafter FN09]{FN09} have constructed a catalog of 
$232$ voids from the Sloan Digital Sky Survey Data Release 5 (SDSS DR5). 
Now that the voids of the FN09 catalog are identified using the HV02 
algorithm, it must provide the most optimal dataset against which 
the PL07 analytic model can be compared.

In this Letter our goal is to constrain the value of $\sigma_{8}$ by 
comparing the extended PL07 analytic model of the redshifted void ellipticity 
function with the observational result from the FN07 catalog of SDSS voids.

\section{REDSHIFTED VOID ELLIPTICITY FUNCTION}

Let us first give a brief overview on the PL07 theoretical model. 
An analytic expression for the probability density distribution of the 
minor-to-major axial ratio, $\nu$, of a void at redshift $z$ on the 
Lagrangian scale $R_{L}$ was found by PL07 as 
\begin{eqnarray}
p(\nu;z, R_{L})
&=& \int_{\nu}^{1}d\mu p[\mu,\nu|\delta =\delta_{v};\sigma(z,R_{L})] \cr
&=& \int_{\nu}^{1}d\mu 
\frac{3375\sqrt{2}}{\sqrt{10\pi}\sigma^{5}_{R_L}}
\exp\left[-\frac{5\delta^{2}_{v}}{2\sigma^{2}_{R_L}} + 
\frac{15\delta_{v}(\lambda_{1}+\lambda_{2})}{2\sigma^{2}_{R_L}}\right] \cr
&&\times\exp\left[-\frac{15(\lambda^{2}_{1}+\lambda_{1}\lambda_{2}+
\lambda^{2}_{2})}{2\sigma^{2}_{R_L}}\right] \cr
&&\times(2\lambda_{1}+\lambda_{2}-\delta_{v})(\lambda_{1}-\lambda_{2})
(\lambda_{1}+2\lambda_{2}-\delta_{v}) \cr
&&\times\frac{4(\delta_{v}-3)^2\mu\nu}{(\mu^{2}+\nu^{2}+1)^{3}}, 
\label{eqn:nu}
\end{eqnarray}
where $\sigma (z,R_{L})$ represents the rms fluctuations of the 
linear density field smoothed on scale $R_{L}$ at redshift $z$, and 
$\{\nu,\mu\}$ (with $\nu \le \mu$) represent the two axial ratios of cosmic 
voids that can be obtained from the inertia momentum tensors of the 
anisotropic spatial positions of void galaxies. The key concept of this 
analytic expression is that the two axial ratios, $\nu$ and $\mu$, 
are related to the largest and second to the largest eigenvalues, 
$\lambda_{1}$ and $\lambda_{2}$, of the tidal field smoothed on the scale 
$R_{L}$ as   
\begin{eqnarray}
\label{eqn:lamu1}
\lambda_{1}(\mu,\nu) &=& \frac{1 + (\delta_{v}- 2)\nu^{2} + 
\mu^{2}}{(\mu^{2} + \nu^{2} + 1)},\\
\label{eqn:lamu2} 
\lambda_{2}(\mu,\nu) &=& \frac{1 + (\delta_{v}- 2)\mu^{2} + \nu^{2}}
{(\mu^{2} + \nu^{2} + 1)},
\end{eqnarray}
where $\delta_{v}$ denotes the critical density contrast of a void linearly 
extrapolated to $z=0$. 
 
PL07 calculated the value of $\delta_{v}$ as the galaxy number density 
contrast as $\delta_{v}\equiv (n_{vg}-\bar{n}_{g})/\bar{n}_{g}$ where 
$n_{vg}$ and $\bar{n}_{g}$ represent the number density of void galaxies 
and the mean number density of all galaxies in a given sample. PL09 found that 
$\delta_{v}\approx 0.9$ on average but also noted a tendency that 
$\delta_{v}$ decreases gradually with the sizes of voids. The Lagrangian 
scale radius $R_{L}$ was calculated as 
$R_{L}\equiv (1+\delta_{v})^{1/3}R_{E}/(1+z)$. Here $R_{E}$ represents the 
effective (comoving) spherical radius of a void defined as 
$4\pi R^{3}_{E}/3 = V$ with the void volume $V$. The values of $\delta_{v}$ 
and $R_{E}$ have to be determined from the observed voids that are to 
be used for comparison. It is worth mentioning here that this relation 
between $R_{L}$ and $R_{E}$ holds good also in redshift space.

Defining the ellipticity of a void as $\varepsilon\equiv 1-\nu$, the 
probability density distribution of the void ellipticities on scale 
$R_{L}$ at redshift $z$ is calculated as 
$p(\varepsilon;z,R_{L})=p(1-\nu;z,R_{L})$.
PL07 originally derived Equations (\ref{eqn:nu}-\ref{eqn:lamu2}) for the 
present epoch $z=0$. It was \citet{LP09} who extended the analytic model to 
higher redshifts \footnote{There are typos in Equations (2) in \citet{LP09}, 
which are corrected here}, according to which the dependence of 
$p(\varepsilon;z,R_{L})$ on $z$ and $R_{L}$ comes from the dependence of 
$\sigma^{2}$ on $z$ and $R_{L}$:
\begin{equation}
\sigma^{2}(z,R_{L})\equiv D^{2}(z)\int_{-\infty}^{\infty}\Delta^{2}(k)
W^{2}(kR_{L})d\ln k, 
\end{equation}
where $W(kR)$ is the top-hat filter of scale radius $R_{L}$, $D(z)$ is the 
linear growth factor normalized as $D(0)=1$, $\Delta^{2}(k)$ is the 
dimensionless linear matter power spectrum. The functional shapes of $D(z)$ 
and $\Delta^{2}(k)$ vary with the background cosmology. Assuming a 
$\Lambda$CDM cosmology, we use the following formula for $D(z)$ and 
$\Delta^{2}(k)$ \citep{lah-etal91,BBKS86} .
\begin{eqnarray}
\label{eqn:lcdmD}
D(z) &\propto& \frac{5}{2}\Omega_{m}
[\Omega_{m}(1+z)^{3}+\Omega_{\Lambda}]^{1/2}\cr
&&\times\int_{z}^{\infty}dz^{\prime}
\frac{1+z^{\prime}}{[\Omega_{m}(1+z^{\prime})^{3} + \Omega_{\Lambda}]^{3/2}}, 
\cr
\Delta^{2}(k) &\propto& \frac{1}{2\pi^{2}}k^{n_s+3}
\left[\frac{\ln(1+2.34q)}{2.34q}\right]^{2}\cr 
\times &&[1 + 3.89q + (16.1q)^{2} + (5.46)^{3} + (6.71q)^{4}]^{-1/2},
\end{eqnarray}
where $q \equiv k/[\Omega_{m}h^{2}{\rm Mpc}^{-1}$] \citep{PD94} and  
$n_{s}$ is a spectral index of the primordial power spectrum. 
The void ellipticity function, $f(\varepsilon;z,R_{L})$, is now defined 
as the differential number density of voids as a function of $\varepsilon$, 
$N_{tv}p(\varepsilon;z,R_{L})$ where $N_{tv}$ is the total number of voids.
\begin{figure} 
\plotone{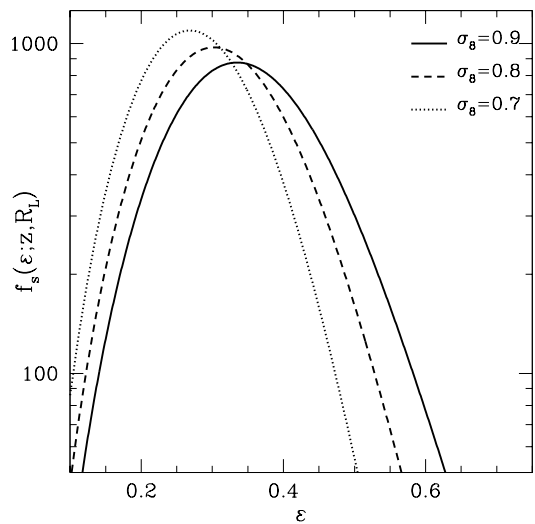}
\caption{Analytic model of the redshifted void ellipticity function for 
the three different cases of the linear power-spectrum normalization 
($\sigma_{8}=0.7,0.8$ and $0.9$ as dotted, dashed and solid lines, 
respectively).}
\label{fig:sig}
\end{figure}
To extend the above analytic expression for the redshifted void ellipticity 
distribution, we consider the simplest case where the redshift distortion 
effect can be described by a single linear distortion parameter, $\beta$, 
which is related to the background cosmology as 
\citep[][and references therein]{ham98}:
\begin{equation}
\beta = \left[\Omega_{m}^{2}+\frac{\Omega_{\Lambda}}{70}
\left(1+\frac{\Omega_{m}}{2}\right)\right]\frac{1}{b_{g}},
\end{equation}
where $b_{g}$ is a linar galaxy bias factor measured in real space. 
Since we are interested in the redshift distortion effect on voids, 
we should use a bias factor of the void galaxies, say $b_{vg}$. 
\citet{bas-etal07} have measured the real-space clustering of the HI 
galaxies and found that the galaxies in low-density regions are 
anti-biased, exhibiting $b_{vg}\approx 0.68$.
Now, the redshifted power spectrum can be approximated at first order as 
\citep{kai87,ham98}
\begin{equation}
\label{eqn:ds}
\Delta_{s}^{2}(k) = 
\left(1+\frac{2}{3}\beta+\frac{1}{5}\beta^{2}\right)\Delta^{2}(k),
\end{equation}
where $\beta$ is calculated using the void galaxy bias factor 
$b_{vg}=0.68$. Replacing the real-space power spectrum $\Delta^{2}(k)$ by 
the redshifted power spectrum $\Delta^{2}_{s}(k)$, we finally obtain an 
analytic expression for the redshifted void ellipticity function, 
$f_{s}(\varepsilon;z,R_{L})$. Figure \ref{fig:sig} plots the 
analytic predictions of the redshifted void ellipticity function for 
the three different cases of the linear power-spectrum normalization 
($\sigma_{8}=0.7,0.8$ and $0.9$ as dotted, dashed and solid, respectively). 
The Lagrangian void scale, redshift, and total number of voids are set at 
the values consistent with the ones used for the void catalog (see \S 3). 
As it can be seen, the void ellipticity function depends very sensitively 
on the $\sigma_{8}$ value.
 
\section{COMPARISON WITH OBSERVATIONAL RESULTS}

FN09 extended the HV02 void-finding algorithm to improve its statistical 
robustness and applied it to the volume-limited spectroscopic sample of the 
galaxies from SDSS DR5 to construct a catalog of $232$ voids 
(available at http://physics.ubishops.ca/sdssvoids). The volume 
limited sample has a total of $52281$ galaxies in a volume of 
$21310400\, h^{-3}{\rm Mpc}^{3}$ (C. Foster in private communication) . 
The mean galaxy number density of this sample is thus 
$\bar{n}_{g}\approx 2.45\times 10^{-3}$ in unit of $h^{3}{\rm Mpc}^{-3}$. 
In the catalog is listed the redshift $z$, effective spherical radius 
$R_{E}$, number of the void galaxies $N_{vg}$, three axis-lengths 
($a,\ b,\ c$ with $a\le b\le c)$ of the best-fit ellipsoids of each SDSS void. 
The effective spherical radius $R_{E}$ of each void is related to its three 
axis-lengths as $R^{3}_{E}=abc$. FN09 determined the best-fit ellipsoid of 
each void with the help of the prescription of \citet{JH01}, assuming a flat 
$\Lambda$CDM cosmology with $\Omega_{m}=0.28$, $\Omega_{\Lambda}=0.72$ 
and $H_{0}=100h$. Throughout this paper, we also adopt the same cosmology, 
setting the other key parameters at $n_{s}=0.96$ and $h=0.71$ from the 
WMAP priors \citep{wmap5}. The mean values of $z$, $R_{E}$, and $N_{vg}$ 
averaged over all voids are found to be 
$\bar{z}=0.114$, $\bar{R}_{E}=24.645\, h^{-1}$Mpc, and $\bar{N}_{vg}=7$, 
respectively.

\begin{figure} 
\plotone{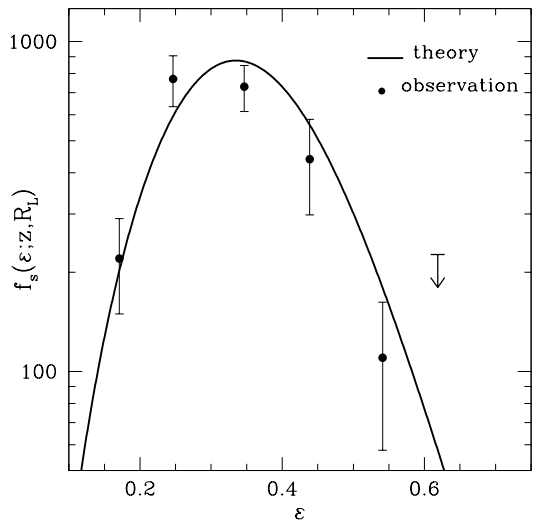}
\caption{Comparison of the observational results (solid dots) with the 
analytic model (solid line) with the best-fit value $\sigma_{8}\approx 0.9$.
The errors include both the cosmic variance and the Poisson noise.The 
downward arrows represent $f_{s}+2\sigma$ for an ellipticity bin where 
$f_{s}<\sigma$.}
\label{fig:com}
\end{figure}

Using information on $a,\ b,\ c$, we measure the ellipticity of each void 
as $\varepsilon=1-c/a$. Binning the ellipticities as $\varepsilon_{i}$, 
counting the number of SDSS voids in each ellipticity bin, and dividing 
the void number counts by $d\varepsilon_{i}$, we determine the observational 
void ellipticity function, $f_{s}(\varepsilon_{i})$. To estimate the 
statistical errors in the measurement of $f_{s}(\varepsilon_{i})$, we separate 
the voids into six subsamples each of which has approximately the same 
number of voids, and then calculate the void ellipticity function separately 
from each subsample. The errors at each ellipticity bin $\varepsilon_{i}$ is 
now calculated as the standard deviation of $f(\varepsilon_{i})$ 
between the six subsamples, which include both the cosmic variance and 
the Poisson noise. Figure \ref{fig:com} plots the resulting void 
ellipticity function as dots. 

The density contrast of each void is computed as 
$\delta_{v}=(n_{vg}-\bar{n}_{g})/\bar{n}_{g}$ where 
$n_{vg}=N_{vg}/V$ with $V=(4\pi abc/3)$. And the mean density contrast 
averaged over all voids is determined to be $\bar{\delta}_{v}\approx -0.97$.
Using the values of $\bar{z}$, $\bar{R}_{E}$ and $\bar{\delta}_{v}$ , 
the Lagrangian void scale is obtained to be $R_{L}\approx 8.534\,h^{-1}$Mpc.
We fit the analytic model of the redshifted void ellipticity function 
to the observational results by adjusting the value of $\sigma_{8}$ with the 
help of the generalized $\chi^{2}$ minimization method where $\chi^{2}$ is 
given as
\begin{equation}
\chi^{2} \equiv  
[f^{O}_{s}(\varepsilon_{i})-f^{T}_{s}(\varepsilon_{i};\sigma_{8})]C^{-1}_{ij}
[f^{O}_{s}(\varepsilon_{j})-f^{T}_{s}(\varepsilon_{j};\sigma_{8})],
\end{equation}
where $f^{O}_{s}(\varepsilon_{i})$ and $f^{T}_{s}(\varepsilon_{i};\sigma_{8})$ 
denote the observed and the theoretical ellipticity function at the 
$i$-th ellipticity bin, respectively. The covariance matrix 
$(C_{ij})$ is defined as $C_{ij}\equiv \langle\Delta f^{O}_{s}(\varepsilon_{i})
\Delta f^{O}_{s}(\varepsilon_{j})\rangle$ where the ensemble average 
is taken over the six subsamples. Since the number of the subsamples 
(six) is larger than the number of the ellipticity bins 
(five), it is guaranteed that $(C_{ij})$ is invertible 
\citep{har-etal07}. The best-fit $\sigma_{8}$ that minimizes $\chi^{2}$ 
is found to be $\sigma_{8}=-0.897\pm 0.037$ where the errors are calculated 
as $\sqrt{2}\left(d^{2}\chi^{2}/d\sigma^{2}_{8}\right)^{-1/2}$.
Figure \ref{fig:com} plots the analytic model with best-fit $\sigma_{8}$, 
demonstrating that the analytic model agrees well with the observational 
results. 
\begin{figure} 
\plotone{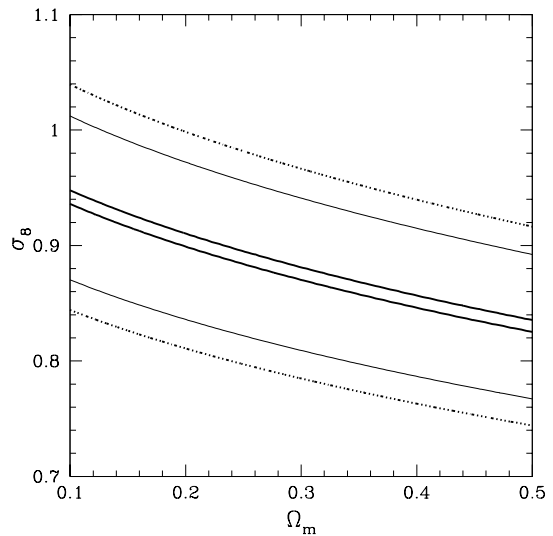}
\caption{Contours of the reduced $\chi^{2}$ ($\chi^{2}_{r}=1,\ 2,\ 3$ 
as thick solid, thin solid and dotted lines, respectively) in the 
$\Omega_{m}-\sigma_{8}$ plane.}
\label{fig:con}
\end{figure}

Although a fixed value, $\Omega_{m}=0.28$, is assumed for the construction 
of the FN09 void catalog, it is worth examining the parameter degeneracy 
between $\sigma_{8}$ and $\Omega_{m}$ when the void ellipticity function 
is used as a probe of $\sigma_{8}$.  Varying the values of $\sigma_{8}$ and 
$\Omega_{m}$ simultaneously, we repeat the whole fitting process. 
Figure \ref{fig:con} plots the contours of $\chi^{2}_{r}=1,\  2,\ 3 $ 
(thick solid, thin solid and dotted lines, respectively) in the 
$\Omega_{m}-\sigma_{8}$ plane, where $\chi^{2}_{r}$ is the reduced 
$\chi^{2}$. As it can be seen, the best-fit $\sigma_{8}$ decreases 
very mildly as $\Omega_{m}$ increases. 

\section{DISCUSSION AND CONCLUSION}

We have constrained the matter power spectrum normalization as 
$\sigma_{8}=0.90\pm 0.04$ by comparing the void ellipticity function from 
SDSS DR5 with the extended PL07 model. It is intriguing to note that our 
result is higher than the WMAP5 value ($\sigma_{8}=0.796\pm 0.036$) but 
consistent with that from the recent work of \citet{LL09} who casted a 
doubt on the accuracy of the cosmological parameters estimated by the 
WMAP team. Using a new improved CMB map constructed by employing an 
independent software scheme, \citet{LL09} have 
found $\sigma_{8}=0.921\pm 0.036$.
 
The advantage of using the void ellipticity function as a probe of 
the power spectrum normalization is that it is purely analytical, free from 
any nuisance parameters and ad-hoc assumptions. Besides, as shown 
recently by \citet{lam-etal09}, the void ellipticity function does not 
depend on the primordial non-Gaussianity parameter, unlike the other 
prominent probe of $\sigma_{8}$, the cluster mass function.
It depends most sensitively on $\sigma_{8}$ among the key cosmological 
parameters and thus it is in principle one of the most optimal probes of the 
power spectrum normalization.  Yet, it is worth noting that 
there is one weak point about using the void ellipticity function as 
a cosmological probe. The number of observable voids at different 
redshifts is relatively small compared with that of observable clusters, 
which means that it tends to suffer from small-number statistics.
The future galaxy surveys may allow us to overcome this limitation.
Our future work is in the direction of forecasting constraints on 
$\sigma_{8}$ and $w$ from the forthcoming galaxy surveys 
by exploiting the void ellipticity function 
(C. Cunha and J. Lee in preparation).

\acknowledgments

I thank C. Foster, C.Cunha and D.Huterer for many inspiring comments. 
I also thank A.E.Evrard and the Physics Department of University 
of Michigan and Michigan Center for Theoretical Physics at Ann Arbor for 
the warm hospitality during my Sabbatical when this research is conducted. 
I acknowledge financial support from the Korea Science and Engineering 
Foundation (KOSEF) grant funded by the Korean Government 
(MOST, NO. R01-2007-000-10246-0).


\begin{thebibliography}{}
\bibitem[Adelman-McCarthy et al.(2007)]{sdssdr5} 
Adelman-McCarthy, J.~K., et al.\ 2007, \apjs, 172, 634 
\bibitem[Bardeen et al.(1986)]{BBKS86}
Bardeen, J. M., Bond, J. R., Kaiser, N., \& Szalay, A. S. 1986, \apj, 304, 15
\bibitem[Basilakos et al.(2007)]{bas-etal07} 
Basilakos, S., Plionis, M., Kova{\v c}, K., \& Voglis, N.\ 2007, 
\mnras, 378, 301 
\bibitem[Colberg et al.(2008)]{col-etal08} 
Colberg, J.~M., et al.\ 2008, \mnras, 387, 933 
\bibitem[Dodelson(2003)]{dod03} 
Dodelson, S.\ 2003, Modern cosmology, (Amsterdam : Academic Press) 
\bibitem[Dunkley et al.(2009)]{wmap5}
Dunkley, J., et al.\ 2009, \apjs, 180, 306
\bibitem[Foster \& Nelson(2009)]{FN09} 
Foster, C., \& Nelson, L.~A.\ 2009, \apj, 699, 1252 
\bibitem[Hamilton(1998)]{ham98} 
Hamilton, A.~J.~S.\ 1998, The Evolving Universe, 231, 185  (astro-ph/9708102)
\bibitem[Hartlap et al.(2007)]{har-etal07} 
Hartlap, J., Simon, P., \& Schneider, P.\ 2007, \aap, 464, 399 
\bibitem[Henry et al.(2009)]{hen-etal09} 
Henry, J.~P., Evrard, A.~E., Hoekstra, H., Babul, A., \& Mahdavi, A.\ 2009, 
\apj, 691, 1307 
\bibitem[Hoyle \& Vogeley(2002)]{HV02} 
Hoyle, F., \& Vogeley, M.~S.\ 2002, \apj, 566, 641 
\bibitem[Huterer \& Ma(2004)]{HM04} 
Huterer, D., \& Ma, C.-P.\ 2004, \apjl, 600, L7 
\bibitem[Jang-Condell \& Hernquist(2001)]{JH01} 
Jang-Condell, H., \& Hernquist, L.\ 2001, \apj, 548, 68 
\bibitem[Kaiser(1987)]{kai87} 
Kaiser, N.\ 1987, \mnras, 227, 1 
\bibitem[Lahav et al.(1991)]{lah-etal91} 
Lahav, O., Lilje, P.~B., Primack, J.~R., \& Rees, M.~J.\ 1991, 
\mnras, 251, 128 
\bibitem[Lee(2006)]{lee06} 
Lee, J.\ 2006, \apj, 643, 724 
\bibitem[Lee \& Park(2009)]{LP09} 
Lee, J., \& Park, D.\ 2009, \apjl, 696, L10 
\bibitem[Lam et al.(2009)]{lam-etal09} 
Lam, T.~Y., Sheth, R.~K., \& Desjacques, V.\ 2009, arXiv:0905.1706 
\bibitem[Liu \& Li(2009)]{LL09} 
Liu, H., \& Li, T.-P.\ 2009, submitted to MNRAS, arXiv:0907.2731 
\bibitem[Park \& Lee(2007)]{PL07} 
Park, D., \& Lee, J.\ 2007, Physical Review Letters, 98, 081301 
\bibitem[Peacock(1999)]{pea99}
Peacock, J. A. 1999, Cosmological Physics, (Cambridge :Cambridge Univ. Press)
\bibitem[Peacock \& Dodds(1994)]{PD94}
Peacock, J. A., \& Dodds, S. J. 1994, \mnras, 267, 1020
\bibitem[Refregier et al.(2002)]{ref-etal02} 
Refregier, A., Rhodes, J., \& Groth, E.~J.\ 2002, \apjl, 572, L131 
\bibitem[Springel et al.(2005)]{spr-etal05} 
Springel, V., et al.\ 2005, \nat, 435, 629 
\bibitem[Tegmark et al.(2006)]{teg-etal06} 
Tegmark, M. et al.\ 2006, \prd, 74, 123507
\bibitem[Van Waerbeke et al.(2000)]{wae-etal00} 
Van Waerbeke, L., et al.\ 2000, \aap, 358, 30 

\end{thebibliography}
\end{document}